A Toy Model for the Cycle Rank Dependence of Stretch at Break in Phantom Chain Network Simulations

Yuichi Masubuchi
Department of Materials Physics, Nagoya University, Nagoya 4649603, Japan
mas@mp.pse.nagoya-u.ac.jp

**Abstract**

The relationship between the topological architecture of polymer networks and their macroscopic rupture remains a fundamental challenge in polymer physics. Recent coarse-grained simulations have revealed that the dependence of stretch at break ($\lambda_b$) on node functionality and reaction conversion can be unified into a universal master curve when plotted against the cycle rank density ($\xi$). However, a theoretical derivation explaining this universality has been lacking. This study proposes a simple mechanical model to describe the $\xi$-dependence of $\lambda_b$. The polymer network is modeled as a mechanical system consisting of a sequence of springs representing localized, highly stretched strands and the surrounding unstretched network. By relating the stiffness contrast between these regions to the network connectivity defined by $\xi$, an analytical expression for the stretch at break is derived: $\lambda_b - 1 \propto (3\xi + 6)/(3\xi + 2)$. The proposed model is validated against phantom chain simulations using both Gaussian and finite extensibility (FENE) springs. The theoretical prediction shows reasonable agreement with simulation data, providing a physical basis for the phenomenological universality observed in polymer network rupture.

**Introduction**

The relationship between the topological architecture of polymer networks and their macroscopic mechanical failure remains one of the most fundamental yet elusive challenges in polymer physics.[1] While the linear elasticity of rubberlike materials is well described by classical theories—such as the affine and phantom network models—the mechanisms governing large-deformation fracture and rupture are far less well understood. A central difficulty lies in identifying the specific structural descriptors that determine fracture limits, particularly in networks containing topological defects or varying node functionalities.

Historically, theoretical approaches to network fracture have relied heavily on idealized assumptions. The classical Kuhn theory [2] estimates the limit of extensibility based solely on the finite extensibility of individual strands, scaling with the square root of the strand degree of polymerization ($N^{1/2}$). Meanwhile, the Lake-Thomas theory [3] connects network structure to fracture energy by assuming that all bonds in a strand must be stressed to rupture. However, these theories often struggle to account for the complex interplay between node functionality ($f$), reaction conversion (p), and the resulting network connectivity. For instance, the effect of node functionality on strength is contentious; while traditional views suggest higher functionality enhances strength[4], recent experiments on Tetra-PEG gels indicate that lower functionality can yield superior fracture properties under certain conditions [5].

To bridge the gap between molecular topology and macroscopic failure, molecular simulations have become indispensable. As recently reviewed [6], the field has evolved from early studies of crack tips [7,8]to large-scale simulations capable of capturing network heterogeneity [9–13]. Despite persistent challenges related to mismatches in spatiotemporal scales between simulations and experiments, and to the validity of coarse-graining in non-equilibrium states, these computational approaches have provided critical insights into how structural defects, such as loops and dangling ends, influence failure.

In a recent series of phantom chain simulations, the rupture of polymer networks formed by end-linking star polymers, with arm number $f$ from the branch point and the end-linking reaction conversion $p$, was systematically investigated [14–21]. By employing a quasi-static energy-minimization approach to eliminate strain-rate artifacts [22–24], the conflicting reports regarding node functionality were resolved. It was demonstrated that the influence of $f$ is highly dependent on the reaction conversion $p$. Specifically, a higher $f$ strengthens the network at low conversions, but this trend reverses at high conversions.

Most significantly, these simulations [15–21] revealed that the complex dependencies of fracture stress and strain on $f$ and $p$ can be unified by a single topological parameter: the cycle rank (ξ). The cycle rank represents the density of independent loops within the network [25–29], and is calculated from the number of network nodes and strands effectively sustaining stress, $n_n$ and $n_{st}$, as follows for the case with $n_n \gg 1$.

$$\xi = \frac{1}{n_n}(n_{st} - n_n) \qquad (1)$$

This value can be calculated for experimental materials for a given set of $f$ and $p$ via mean-field theories [30,31]. Simulation data for strain $\lambda_b$, stress $\sigma_b$, and work $W_b$ of rupture were observed to collapse onto universal master curves when plotted against the cycle rank density.

However, while these simulation results establish a strong phenomenological correlation between cycle rank and fracture limits, a theoretical derivation explaining why fracture strain is governed by cycle rank is notably absent. In this paper, a theoretical model is proposed to describe the dependence of fracture strain on cycle rank in polymer networks. By reconsidering the distribution of load and network connectivity through the lens of cycle rank, this study aims to provide a physical basis for the universality observed in simulations.

**Model**

Figure 1 schematically exhibits the model of this study. As seen in the leftmost panel (a), at the network rupture, a highly stretched region is observed, while most of the system remains rather unstretched. Inspired by this observation, the entire system is replaced by a sequence of three springs with different stiffnesses, $k_s$ and $k_h$, as shown in panel (b). Here, $k_s < k_h$ is assumed, that the middle spring is primarily stretched, and rupture occurs when the stretch of the middle spring exceeds a critical value $\lambda_{bs}$. The springs are assumed to be Hookean throughout the analysis. For the spring sequence, the spring constant of the entire system, $k$, is written as follows.

$$\frac{1}{k} = \frac{2}{k_h} + \frac{1}{k_s} \qquad (2)$$

Concerning the relation between $k_s$ and $k_h$, let us consider the configuration in panel (c), which naively represents the boundary between stretched and unstretched regions. Namely, for each node, a single strand in the stretched region (the middle spring) balances the force with the other strands in the unstretched region.

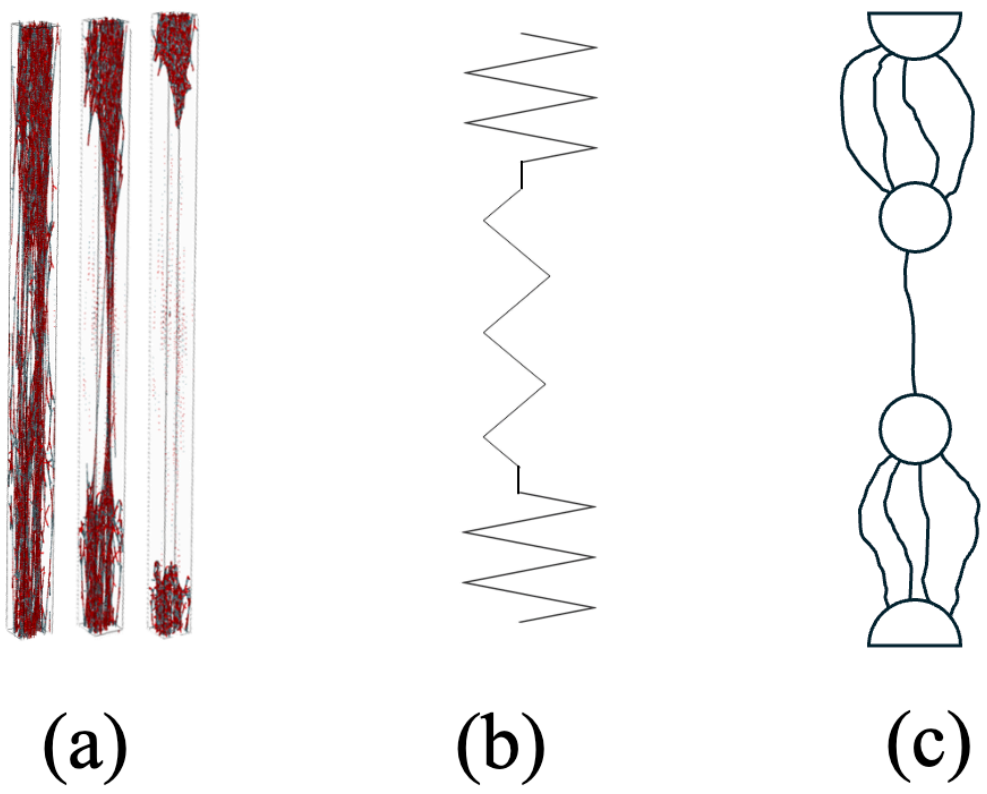

**Figure 1** Typical simulation snapshots at the network rupture (a), the sequential 3-spring model (b), and the naïve strand representation of the 3-spring model (c).

With the number of contributing strands for single nodes at both ends $\nu_s$, $k_h$ is written as below.

$$k_h = \nu_s k_s \qquad (3)$$

Let us consider the stretched network with the stretch ratio $\lambda$ defined as $\lambda = L/L_e$. Here, $L$ is the length of the system along the stretching direction, and $L_e$ is before stretch. Assuming that the stretch of the middle spring reaches $\lambda_s$ under a given exerted tension $F = k_s(\lambda_s - 1)$, the entire stretch $\lambda$ is written as below from eqs 2 and 3.

$$\lambda - 1 = \frac{F}{k} = \frac{F}{k_s}\left(\frac{2}{\nu_s} + 1\right) \qquad (4)$$

$\nu_s$ can be related to the cycle rank $\xi$ defined in eq 1. In the construction of Fig. 1 (c), $n_{st}$ can be written as follows, since the system contains $\nu_s$ strands at both ends, a single strand at the middle, and three nodes.

$$n_{st} = \frac{n_n}{3}(2\nu_s + 1) \qquad (5)$$

Equations 1, 4, and 5 give the relation between the cycle rank $\xi$ and the stretch at break of the entire system $\lambda_b$, when $\lambda_s$ reaches its critical local stretch for break $\lambda_{bs}$.

$$\lambda_b - 1 = (\lambda_{bs} - 1)\left(\frac{3\xi + 6}{3\xi + 2}\right) \qquad (6)$$

Note that the proposed three-spring model is not a literal claim about three distinct regions in the network, but rather a coarse-grained representation of the weakest-link rupture: at break, stress concentrates at a single strand (the "middle" element), while the surrounding network acts as a load-distributing scaffold (the "outer" elements). Equation 6 corresponds to the naive three-element decomposition and expresses a simple topology to capture the competition between series-bottleneck and parallel-bundle, although different constructions can be considered. Thus, the parameter $\lambda_{bs}$ is a parameter that describes the properties of the weakest link in the coarse-grained picture, and may not correspond to the critical extension of breakage for the single strand. Hereafter, $\lambda_{bs}$ is regarded as a fitting parameter.

**Simulations**

The theoretical prediction for $\lambda_b$ is evaluated by comparing it with phantom network simulations. Since the simulation data have already been published separately [14–21], a brief explanation of the simulation is given here. Rouse-Ham-type star-branched prepolymers with $f$ branching arms, each consisting of $N_a$ beads connected by linear springs, are dispersed in a simulation box with periodic boundary conditions. Units of length, energy, and time are chosen as the equilibrium bond length $a$, the thermal energy $k_BT$, and the bead diffusion time $\tau = \zeta a^2/k_BT$, where $\zeta$ is the friction coefficient of the single bead. The bead number density is fixed at 8, and the number of branching arms in the system is ca. 3300. After the system equilibrates, the end-linking reactions are turned on. The reaction occurs only between different prepolymers to prevent the formation of primary loops, as in the experimental setup for PEG gels [32]. In the resultant networks, the spring number per strand is $N_s = 2N_a + 1$. The simulation snapshots are stored at designated conversion $p$, ranging from 0.6 to 0.95. The stored networks are repeatedly subjected to a set of energy-minimization (without Brownian motion) and infinitesimal stepwise deformations. During this elongation, each bead position is traced, although further smeared simulations are possible by replacing connected bonds with a single spring. Nevertheless, each bond length is monitored, and bonds are removed when their length exceeds the designated critical length $b_{\mathrm{c}}$. The total energy for the Gaussian network case is as follows.

$$U = \frac{3k_BT}{2a^2}\sum_{i,j}\mathbf{b}_{ij}^2 \qquad (7)$$

Here, $\mathbf{b}_{ik}$ is the bond vector between beads *i* and *j*. For this case, $b_{\mathrm{c}}$ was chosen at $\sqrt{1.5}$~1.22. For comparison, FENE networks are also examined with the total energy given below.

$$U = -\frac{3k_BTb_{\max}^2}{2a^2}\sum_{i,k}\ln\left(1-\frac{\mathbf{b}_{ik}^2}{b_{\max}^2}\right) \qquad (8)$$

Here, $b_{\max}$ is a nonlinear parameter set to 2, and $b_{\mathrm{c}}$ was chosen to be $\sqrt{3.6}$~1.89. The stretch at break, $\lambda_b$, is determined from the peak of the nominal stress-strain curve. The cycle rank $\xi$ is calculated according to eq 1, which varies depending on $f$ and $p$, and is also slightly different among simulation runs started from different initial configurations. For statistics, eight independent simulations are performed for each condition.

**Results and Discussion**

Figure 2 compares eq 6 with the previous simulation results for various strand molecular weights [20]. By varying the parameter $\lambda_{bs}$, eq 6 agrees with the data within the error bars.

The presented $\xi$-dependence of $\lambda_b$ has been described by a power law as $\lambda_b \propto \xi^{-\alpha}$ [15–21,33]. However, such a description is unphysical in the limit of $\xi \to \infty$, where $\lambda_b$ diminishes to zero. For well-developed networks with a large $\xi$, $\lambda_b$ is expected to approach a specific value, rather than zero. Eq 6 is consistent with this expectation, although in the examined range of $\xi$, such a behavior is not observed.

In the limit of $\xi \to 0$, eq 6 suggests a constant value. The data are inconsistent with this prediction, as they show a power-law-like increase. This discrepancy arises because, with a small value of $\xi$, the model setting shown in Fig 1 is incorrect. Namely, the theory assumes only one single strand is stretched, whereas in this $\xi$ range, a couple of single strands are sequentially connected and stretched to realize large $\lambda_b$ values. Indeed, the data at $\xi \sim 0.02$ correspond to $f = 3$ and $p = 0.6$, which yield fewer than 2 arms per reacted branch point.

To take into account such an effect, eq 2 should be modified as

$$\frac{1}{k} = \frac{2}{k_h} + \frac{M_{ss}(\xi)}{k_s} \qquad (9)$$

Here, $M_{ss}(\xi)$ is the number of stretched strands dependent on $\xi$. $M_{ss}$ is probably related to the number of strand-extending prepolymers discussed previously[14,15]. Nevertheless, for simplicity, this study employs $M_{ss} = 1$, given that eq 6 reasonably reproduces the data in most of the examined range of $\xi$. The other reason for the discrepancy in the small-$\xi$ range is that the theory does not account for nonlinearity, which arises even under large deformations in networks of linear springs.

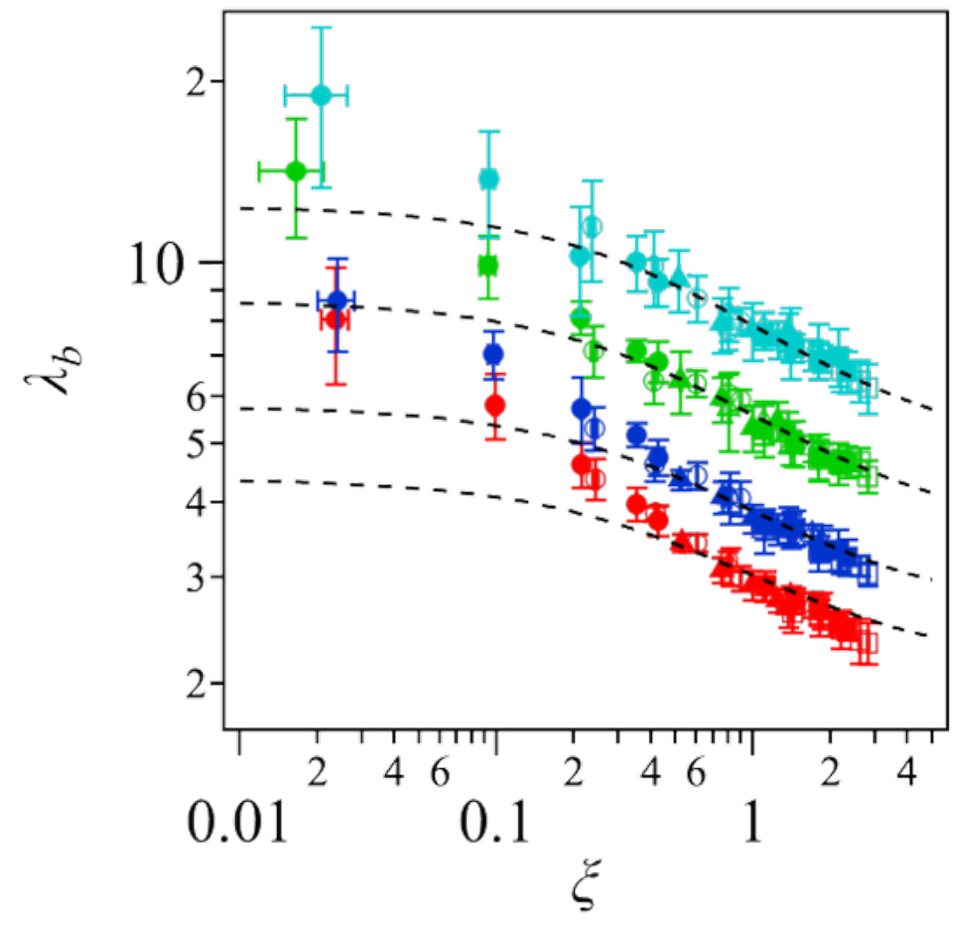

**Figure 2** Comparison of the theory (eq 6) with the simulation results for Gaussian spring networks [20] with the strand length at 7 (red), 11 (blue), 21 (green), and 41 (sky blue). Symbols indicate $f$ = 3 (filled circle), 4 (unfilled circle), 5 (filled triangle), 6 (unfilled triangle), 7 (filled square), and 8 (unfilled square). Error bars correspond to the standard deviations for eight different simulation runs for each condition. Broken curves show eq 6 with varying $\lambda_{bs}$.

Although the basic setting differs from the theoretical model, eq 6 is compared with simulation results using FENE springs [21] in Fig 3. The other parameters are the same as those for the Gaussian networks, except for the bond-breaking parameter $b_c$, which was $\sqrt{1.5}$ for the Gaussian networks and $\sqrt{3.6}$ for the FENE networks, as mentioned above. The results are essentially the same as those shown in Fig 2, demonstrating that the proposed equation reasonably describes the data for $\xi \gtrsim 0.02$.

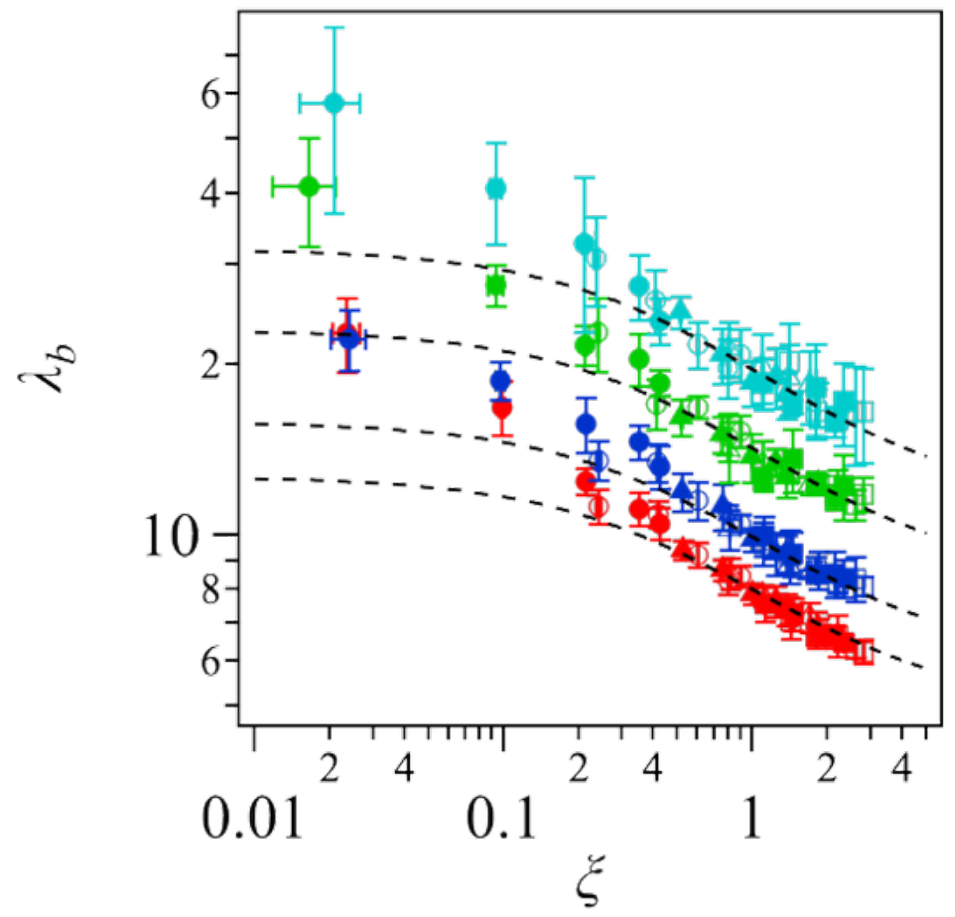

**Figure 3** Comparison of the theory (eq 6) with the simulation results for FENE spring networks [21]with the strand length at 7 (red), 11 (blue), 21 (green), and 41 (sky blue). Symbols indicate $f$ = 3 (filled circle), 4 (unfilled circle), 5 (filled triangle), 6 (unfilled triangle), 7 (filled square), and 8 (unfilled square). Error bars correspond to the standard deviations for eight different simulation runs for each condition. Broken curves show eq 6 with varying $\lambda_{bs}$.

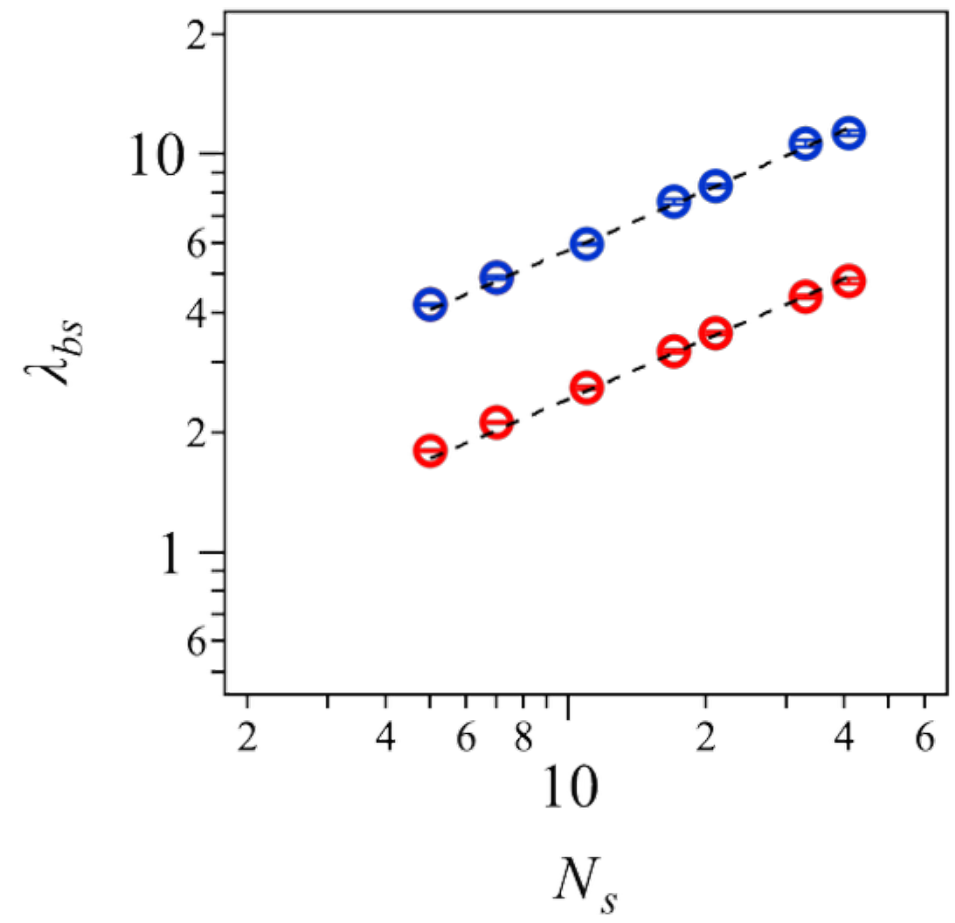

**Figure 4** $\lambda_{bs}$ obtained from the fitting shown in Figs 2 and 3 plotted against $N_s$. Red and blue symbols indicate the cases in Figs 2 (Gaussian springs) and 3 (FENE springs), respectively.

From the fitting in Figs 2 and 3, the value of $\lambda_{bs}$ is extracted and plotted against the strand molecular weight $N_s$ in Fig 4. The dotted lines show a slope of 0.5, which is expected from the model construction. The difference in the absolute value is due to the chosen bond-breaking parameter.

**Concluding remarks**

According to the crude mechanical model, the $\xi$-dependence of $\lambda_b$ for phantom chain networks was derived as $\lambda_b - 1 = (\lambda_{bs} - 1)(3\xi + 6)/(3\xi + 2)$, where $\lambda_{bs}$ is the fitting parameter. The proposed equation was tested against reported simulation results for Gaussian and FENE spring networks, demonstrating reasonable agreement for $\xi \gtrsim 0.2$. $\lambda_{bs}$ is in proportion to the square root of the strand molecular weight $N_s$.

Although the proposed function is promising for further testing, several issues remain open. For instance, the relation between $\xi$ and the model spring constants is rather arbitrarily chosen. In addition, the proposed theory addresses only $\lambda_b$ and does not describe stress at break or work of fracture. The theory should be extended to cover such problems.

The other, further critical issue is the comparison to experiments. Data on network rupture with controlled $f$ and $p$ are rare due to experimental difficulties in precisely capturing rupture. A recent study [34] attempted to evaluate $\lambda_b$ of polyethylene glycol networks, and the resulting $\xi$-$\lambda_b$ relation is close to that discussed in this study. However, they employed double-network gels, in which the PEG networks are supported by the other tough network; thus, the results include the effect of the second network, and direct comparison to the proposed theory is not straightforward.

Nevertheless, supplemental studies are ongoing, and the results will be published elsewhere.

**Acknowledgements**

This study was partly supported by the Hibi Foundation and JSPS KAKENHI (26H02291)